# INVESTIGATING THE RESPONSE OF SINGLE PAD AND HYBRID PIXEL DETECTORS (Si, GaAs, CdTe, CZT) FOR 226Ra AND 137Cs

**G. Chelkov[1], S. Abou El-Azm[1], U. Kruchonak[1,2] , T.H.B. Phi[1,3] , G. Lavrov[1,4], V. Rozhkov[1], R. Sotenskii[1] , A. Lapkin[1]**

*[1]JINR, Dubna, Russia, [2]INP, Minsk, Belarus, [3]Institute of Physics, Vietnam Academy of Science and Technology, 10 Dao Tan, Ba Dinh, Hanoi, Vietnam, [4]Dubna State University, Dubna, Russia*

[1]Corresponding author E-mail: said@jinr.ru

## Abstract:

The development of hybrid pixel detectors for particle detection with high spatial resolution in high energy physics experiments has spun off a number of developments with applications in imaging, especially biomedical imaging, and also imaging in X-ray astronomy and medical application. The energy resolution of pixel detector and cluster size in depends on applied voltage and compression between pixel detector and pad detector have no may attention especially in case of pixel detector .

The purpose of this work was to compare the characteristics of using pixel and single pad detectors made of three semiconductor materials (Si, GaAs and CdTe) under the same conditions using Alpha and gamma rays.

This paper studies more details the energy resolution, cluster size and signal magnitude of three type of pixel detectors (Si, GaAs:Cr, CdTe) and compression with three pad detectors (Si, GaAs, CdTe) in dependence on applied voltage using 226Ra and 137Cs, the results of dependence of resolution, cluster size and signal magnitude on applied voltage are presented for pixel detectors and pad detectors.

## Introduction :

Hybrid pixel detectors are used in various fields of physics experiments and applications. In recent years, they have shown great impact in the field of biology, medicine, space, industry, education and geology[1]. In high energy physics, these detectors are used for particle registration and especially for tracking and timing application[2]. There are many work related to Si has been used as a sensor material in such detectors, however it has some disadvantages due to its relatively low attenuation coefficient for X-ray, a disadvantage that paved the way for other sensor materials such as GaAs, CdTe, CdZnTe can be used as a sensor for pixel detector, they have some advantage better that Si because have high efficient registration in compression with Si sensor. There is no any work related to compare between pad detector and pixel detector with different sensors at the same condition’s, in order to explain the advantage or disadvantage of these detector, it will be important to compare the pixel detector with pad detector, but first of all it is possible to give short information about principle work of pad detector and pixel detector. A semiconductor detector is a radiation detector based on a semiconductor, such as silicon, GaAs, CdTe ....., to measure the effect of incident charged particles or photons. In general, semiconductors are inorganic or organic materials that can control their conduction depending on their chemical structure, temperature, illumination, and the presence of dopants. The energy required to produce electron-hole pairs is very low. In semiconductor detectors, the statistical variation of the pulse height is smaller and the energy resolution is higher.

In the case of the pixel detector, the indicated principle holds for semiconductor detectors and applies both pixel and single-pad devices is based on the direct conversion of radiation into charge in a sensitive layer. It can be electronically collected, detected, and processed by imaging. For example, the absorption of X-rays in a silicon sensor leads to the generation of electrons and hole pairs in the sensor with a charge proportional to the X-ray energy. The applied electric field allows fast signal collection with small signal scattering. Therefore, high sensitivity and spatial resolution can be achieved.

The readout principle is based on pixel imaging. Each sensor pixel is connected to an ASIC (3) pixel by a miniature bump bond. The basic architecture of each ASIC pixel consists of a signal amplifier, a discriminator and a counter. The threshold can be controlled by the user via software and only incoming pulses above the threshold are counted. Digital detection provides noise-free detection of X-rays because of the energy threshold.

This paper presents the results of signal magnitude and energy resolution of four pad detectors (Si, GaAs, CdTe and CdZnTe) and pixel detectors (Si, GaAs, CdTe)[4,5] for the detection of alpha and gamma particles and their dependence on the applied voltage.

**Experiment :**

In this experiment, four types of pad detectors made of Si, GaAs, CdTe and CdZnTe as shown in Table (1) and three types of pixel detectors (Table 2) are presented:

**Table 1:** types of pad detectors made of Si, GaAs, CdTe and CdZnTe

| **Materials** | **Thickness, µm** | **Size, mm** | **Type (producer)** | **Bias volt,v** |
|---|---|---|---|---|
| Si | 400 | 5×10 | n-type Si, 2 sensors assembly | 2,5,10,50,100,200 |
| GaAs:Cr | 250 | 5×5 | High resistive GaAs, compensated by Chromium | 100,250,400,500 |
| CdTe | 1000 | 4×4 | Schottky type (Acrorad) | 100,250,500,750,1000 |
| CdZnTe | 1000 | 5×5 | (Crystal Nord) | 100,250,500,750,1000 |

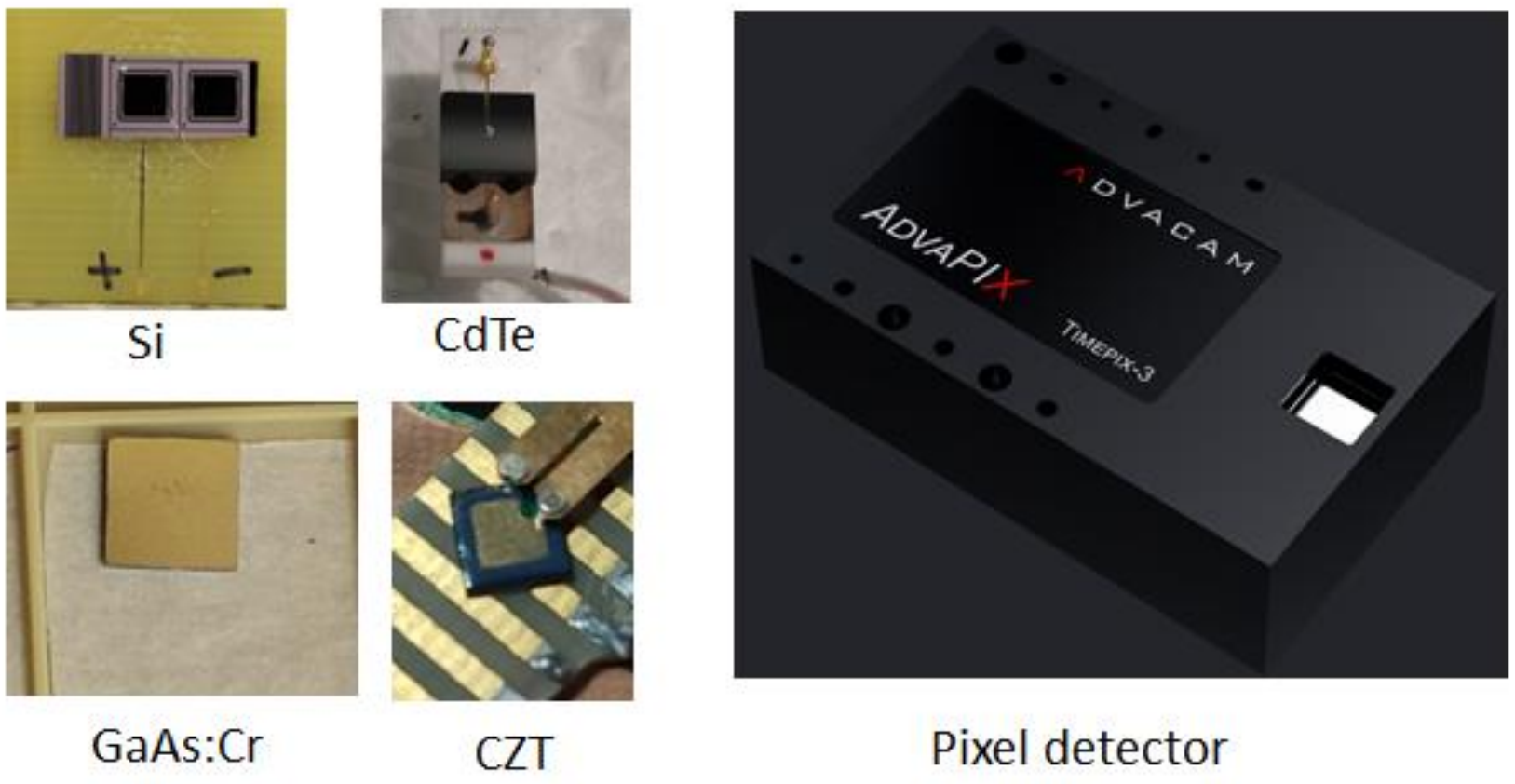

Figure 1: Single pad detectors and pixel detector

**Table 2:** types of detectors  Si, GaAs, CdTe

| Material | Thickness, µm | Size, mm | Pixel matrix | Pixel size, µm | Type | Bias volt,v |
|---|---|---|---|---|---|---|
| Si | 300 | 14.08×14.08 | 256×256 | 55 | Timepix3 | 25,50,75 ,100,200 |
| GaAs:Cr | 500 | 14.08×14.08 | 256×256 | 55 | Timepix3 | 100,200, 250,500, 750 |
| CdTe | 1000 | 14.08×14.08 | 256×256 | 55 | Timepix3 | 100,200, 250,500, 750 |

For the pad detector, the signal was measured using the next setup as shown in Figure (2) which explain the diagram of the signal measurement setup , these measurement's set up was published in work[6]. The collimated source of 226Ra α-particles or 137Cs γ -source passes over the biased sensor and produces charge carriers by ionization. The generated charge carriers are collected and the resulting signal is read out by a charge sensitive amplifier and then digitized by the DRS4 Evaluation Board[7,8]. The data is read out via USB to a computer for processing. For the pixel detector, the Pixet software was used in time-over-threshold (TOT) mode[9] to measure the energy of α and y particles. In TOT mode, the signal amplitude is measured indirectly by measuring the time the signal is above the threshold by counting pulses from a built-in generator[10]. The generator frequency was changed from 10 MHz to 25 MHz. All detectors were measured in the air without vacuum and the temperature was stabilized at 180 $^{0}$C for all measurements, the radioactive α-source was putted directly on the sensor after removing the cover of detector.

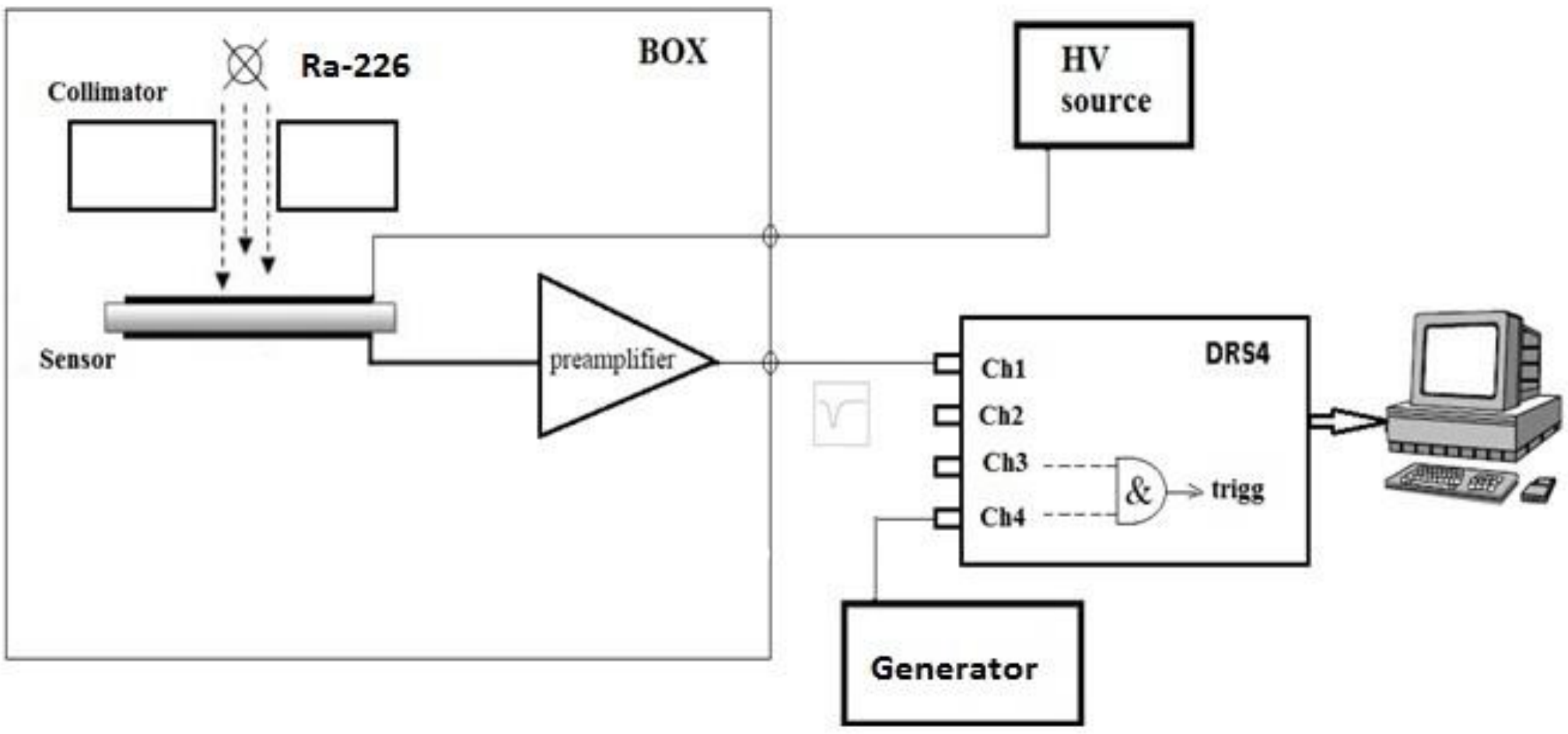

Figure 2: Block diagram for single pad sensor measurement.

**Results and discussions:**

**Section 1 (226Ra):**

The 226Ra spectra for Si pad detector and Si pixel detector are presented in figure (3a). The applied voltage is 50 V for both detectors, the thickness for Si pad detector is 400μ and 300μ for Si pixel detector.

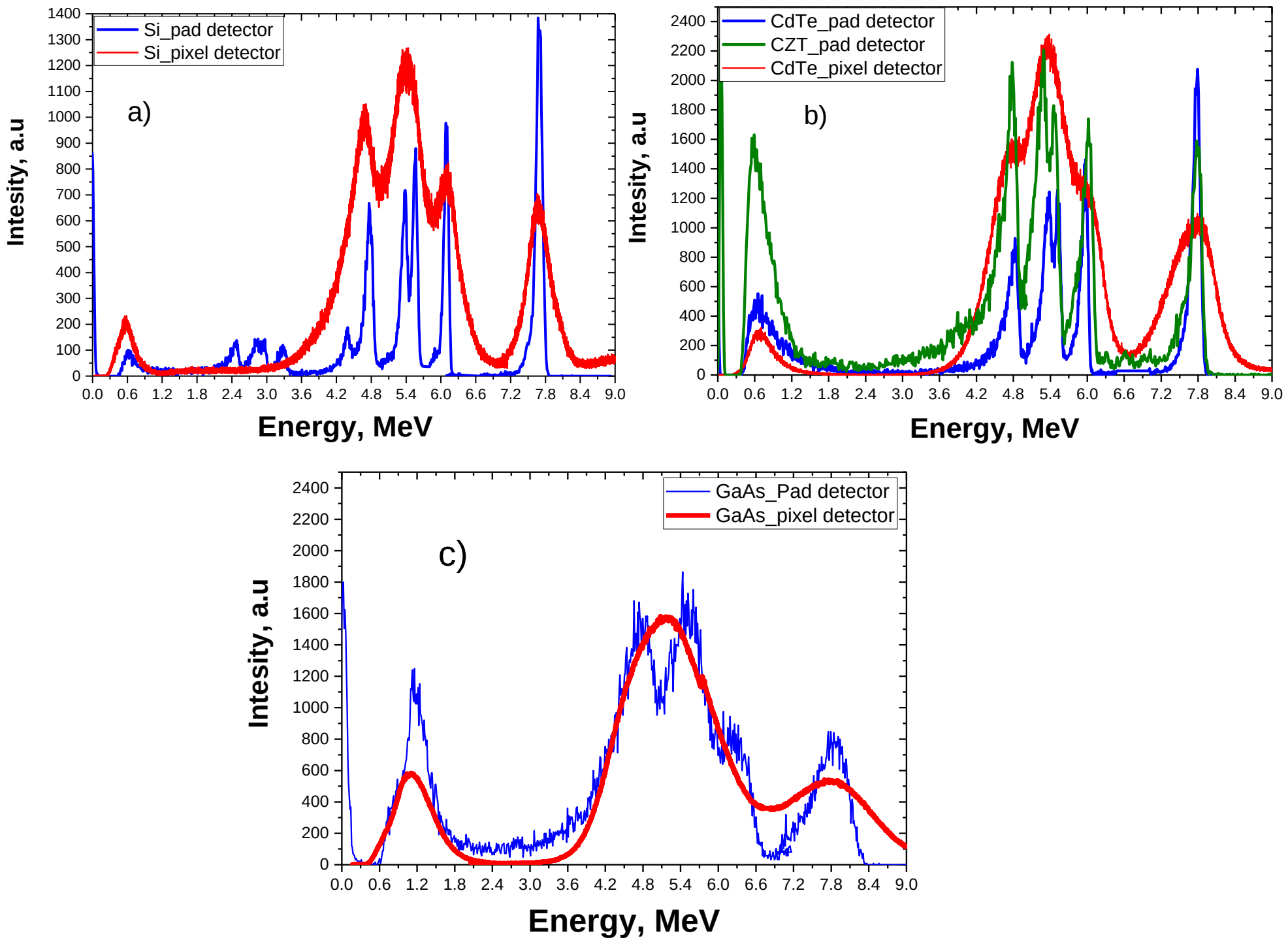

Figure 3 (a,b,c) 226Ra spectra for Si, CdTe, CZT, GaAs pad and pixel detectors respectively.

Per-pixel calibration described in the literature [11] was not performed for pad and pixel detectors. This type of calibration has proven itself well for X-ray imaging, where photons with energies greater than 200 keV are not used. In our work, the energy range was up to 8 MeV. Therefore, a calibration based on knowledge of the energy line values in the decay spectrum of 226Ra (7.6 MeV) and 137Cs (662 keV) sources was used. Comparison of the spectral line values of known radioactive sources and the spectra we obtained allowed us to obtain a one-to-one correspondence between the energy scale and the detector channel values in case of pad detector and TOT value for pixel detector .

As well known the 226Ra have five peaks as following (the energy lines from the decay chain of 226 radium (226 radium - 4784 keV, 210 polonium - 5304 keV, 222 radon - 5490 keV, 218 polonium - 6002 keV, 214 polonium - 7687 keV) ) [12]. As can be seen from these figure 3 (a,b,c), all five α-peaks of are separated in case of Si pad detector and pixel detector, the energy resolution of peak (7.6MeV) is better for pad detector, it is about 1% and about 5% for the Si pixel detector.

Figure (3b) shows the spectrum of 226 radium obtained using pad detectors with a sensor made of CdTe and CZT, as well as a CdTe pixel detector. The thickness of the sensor of each detector is 1000 μm. The bias voltage for each detector was the same and was 750 V. As can be seen,. For these detectors, the energy resolution is ~about 1%. For a pixel detector, the peaks merge and the energy resolution is ~10%)

Figure (3c) shows the emission spectrum of 226 radium by pad and pixel detector with GaAs sensor material. The bias voltage is 750 V. As can be seen from the figure, all peaks of 226Ra are separated for the pad detector. However, for the pixel detector, there is overlapping between peaks (4.7, 5.3, 5.4 and 6 MeV) , these may be related in case of GaAs , the charge collection only electron , also from figures 3 (a,b,c) there is a peak at low energy around (1 MeV), these peaks can be associated with the spectra of beta and alpha particles of lower energies. However, the largest contribution to this peak is noise arising from clustering errors in the registration of events.

**Bias voltage effect on cluster size :**

When a particle goes through the detector, the charge collected can be spread due to diffusion over several pixels. If there is enough charge in order to pass the threshold then these neighboring pixels are grouped together to form a cluster corresponding to the passage of one particle. The cluster size corresponds to the number of pixels contained in the cluster. The results for the cluster size as a function of bias voltage presented in figure 4(a,b) of the three sensors (Si, GaAs, CdTe) pixel detectors.

A general conclusion is that due to shorter distance for diffusion, there is less charge sharing in a thin sensor than in more thickness sensor. Another explanation could be the different charge carriers for the different assemblies (electrons diffuse more than holes thanks to larger diffusion coefficient mobility for electrons in sensors). The thickness of the depletion region and the magnitude of the electric field both increase with the increase in bias voltage. This leads to two competing effects. The first effect is the increase in charge collection due to larger depletion zone and stronger electric field. More charge collection means it is easier to pass the threshold, means more pixels are more excited. This explains the initial rise in clusters as a function of bias voltage. The second effect is the decrease in drift time when the bias voltage is increased. Shorter drift time means less diffusion which leads to less charge sharing and on average, smaller clusters as, these as an explanation for decreasing the cluster size with increase of bias voltage.

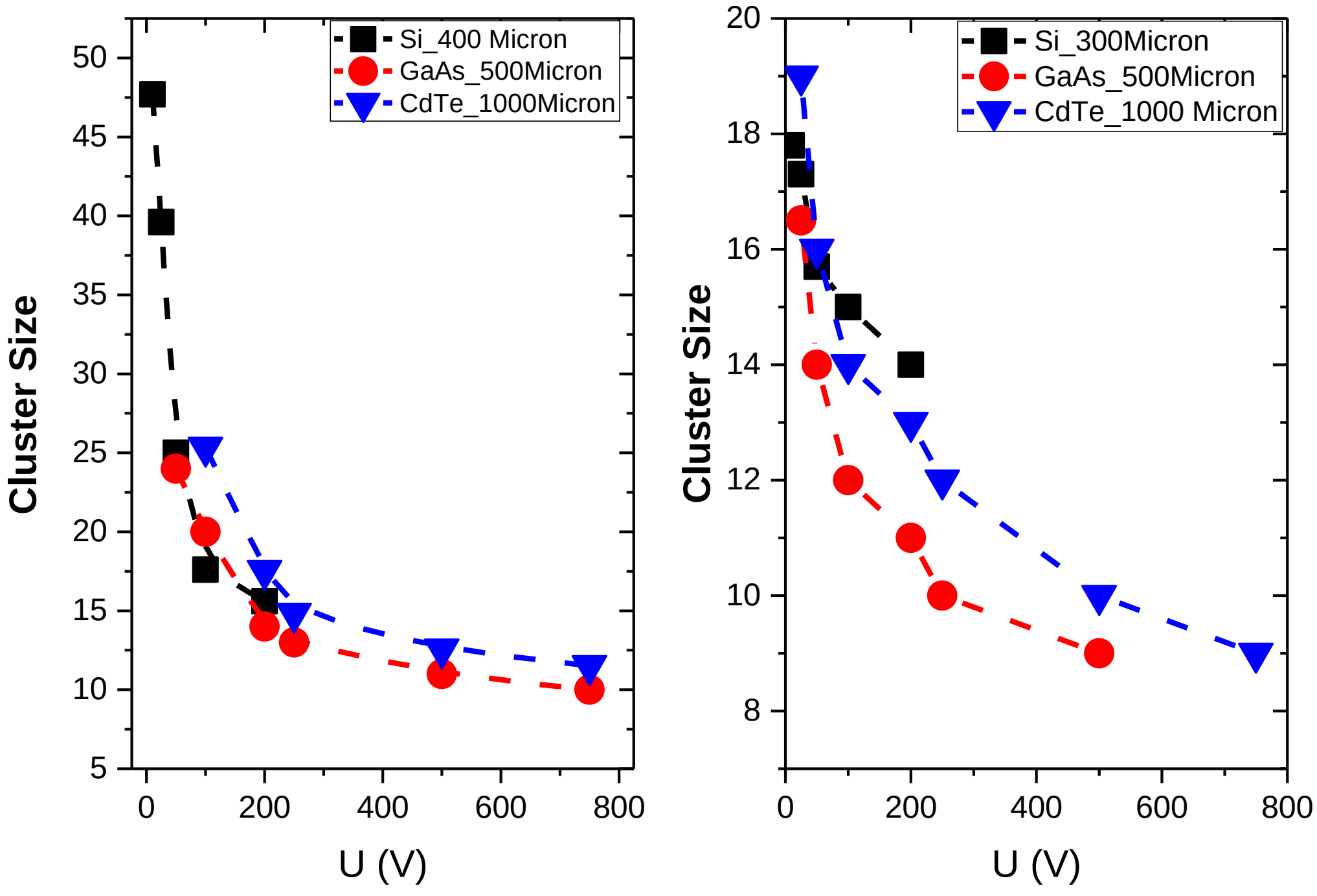

Figure (4) Cluster size vis applied voltage, a) 226Ra, b) 137Cs for pixel detector.

**Bias voltage effect on energy collection (signal) for Alpha particle's:**

The energy collection in the detector is related to the TOT measurement. In order to extract the energy from the TOT measurement must perform calibration of the detector especially in case of pixel detector , as we mentioned before how to convert TOT to energy.

Figure 5 (a,b) show the dependence of the energy collection (TOT) on the bias voltage. These results show that the collected energy increases when the bias voltage is increased. The energy  is usually due to charge being released before the particle enters the depletion region, then the charge does not drift. Energy can be lost also due to recombination in the depletion area or diffusion of the charge to neighboring pixels that does not collect enough charge in order to pass the threshold .

Stronger electric field and thicker depletion region, which are both the result of higher bias voltage, lower the chances of these kinds of energy loss. That explains the increase in energy collected. After reaching a certain bias voltage (depletion voltage), above which the energy collection approximately remains constant, the sensor is fully depleted i.e. the depletion region thickness equals the sensor thickness (almost all free charge carries are collected by electrodes) in next places where depleted mention must be fixed too.

The energy collection graph has two regions. One is for low bias voltages, for which the electric field has not yet reached the edge of the pixel. The second region is for bias voltage lower then depletion voltage but still higher then the voltage necessary for the electric field to reach the edge of the pixel and the bias voltages higher then depletion voltage, for which the energy collection is approximately constant.

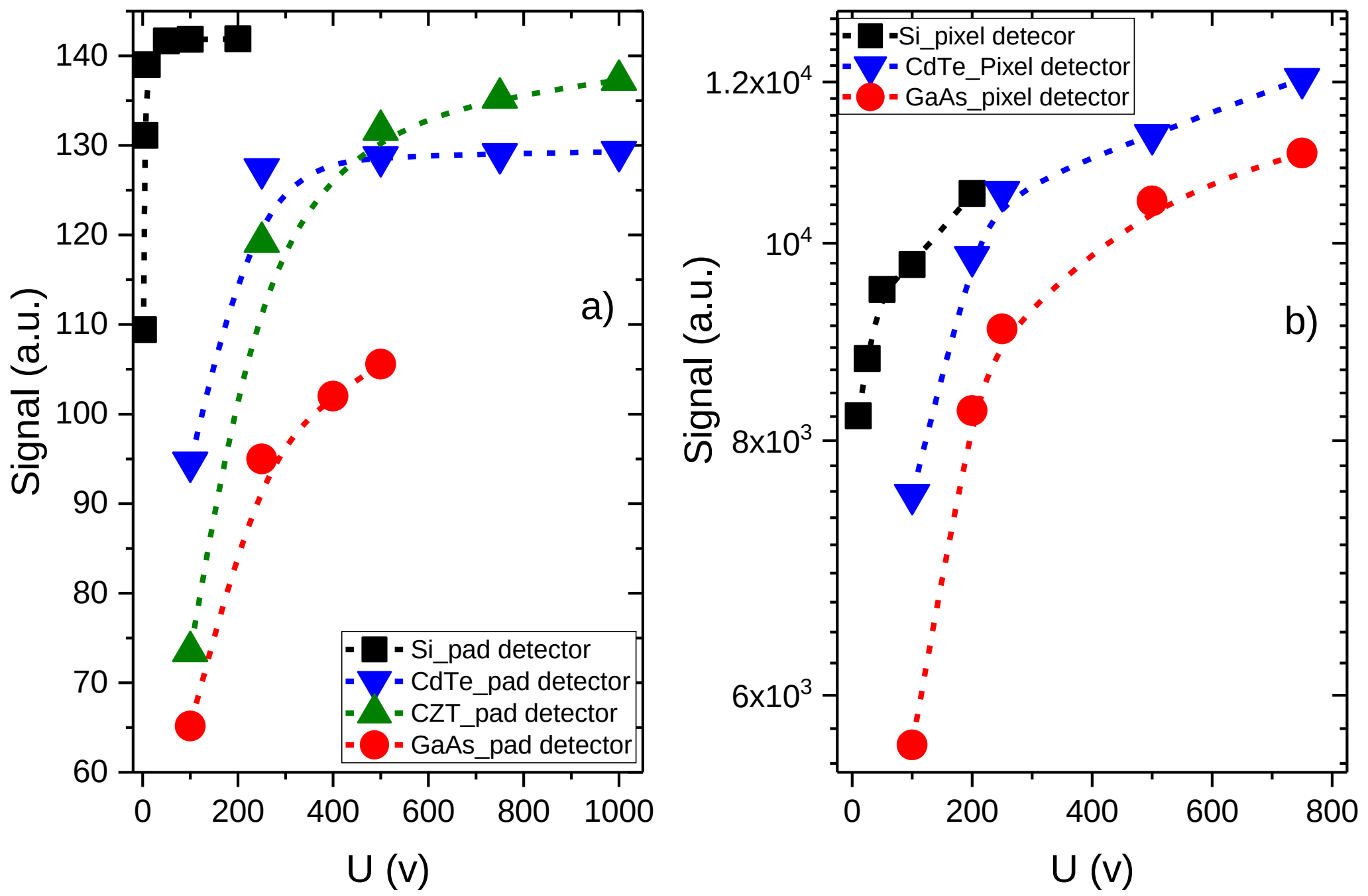

Figure (5) Signal in peak (7.6 MeV) 226Ra for Si, CdTe, CZT, GaAs:Cr pad (a) and pixel (b) detectors.

The relation between signal magnitude in arbitrary units for pad and TOT for pixel detectors for different bias voltages re presented in figure 5(a,b). In figure (5a) the increasing of applied voltage leads to increase the signal for all sensors and then reach to saturation for pad detectors and pixel detector as expected. From the figure (5a) the maximum signal in Si is higher and reach saturation with applied voltage 50V. This indicates that the detector is fully depleted in contrast to CdZnTe and in GaAs. Also the signal higher because the energy of $e^-$-h pair production in Si is 3.6 eV versus 4.46 eV, 4.6 eV and 4.2 eV for CdTe, GdZnTe and GaAs respectively. In CdTe and CdZnTe the signal is practically same as presented in figure (5a). The lower signal in GaAs detector can be explained that it still did not reach saturation by applied voltage, also the poor resolution indicates incomplete charge collection, as well as possibly a large thickness of the dead zone in GaAs detector. The same situation in case of pixel detector as shown in figure (5b) for Si , CdTe and GaAs , the signal is increase and then reach to saturation , the reasons of the saturation state was before in case of pixel and pad detector .

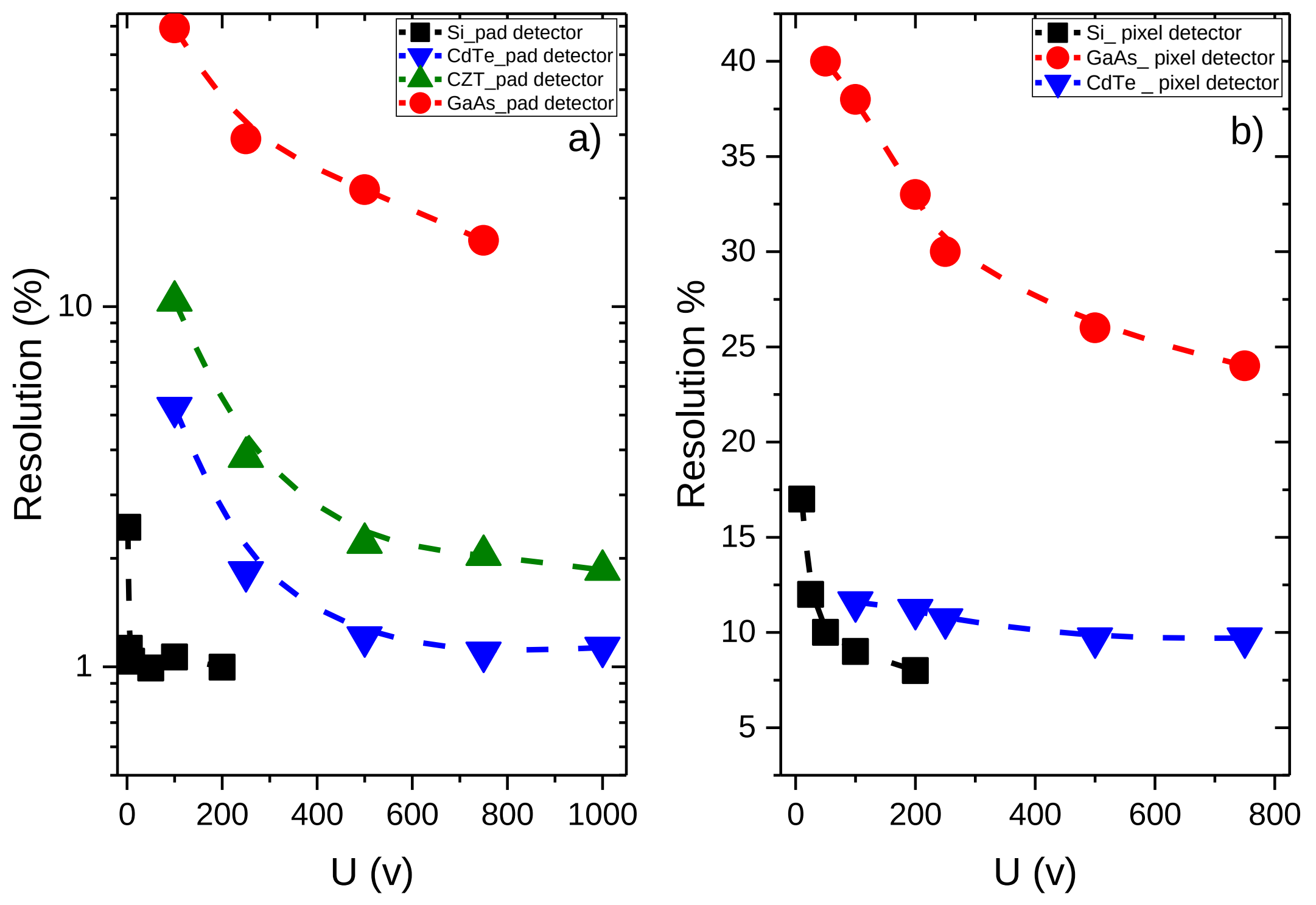

Figure (6) Resolution versus applied voltage (a) pad detector
( b) pixel detector using 226Ra

**Bias voltage effect on detector resolution for Alpha particle's:**

The energetic resolution is important for any detector, for 226Ra have five peaks , in our case we used only the peak at (7.67 MeV) for compassion the energy resolution for different sensors , in figure (5a) presents the resolution of peak (7. 67 MeV) depend on the bias volt for 226Ra, as can be seen from the figure the resolution becomes better with increasing applied volt, as expected, up to applied voltage 200 the resolution of Si pad detector is about (1%) and for CdTe pad detector up to 750V the resolution is about (1%) and for CZT pad detector at 750V the resolution is about (2%), in case of GaAs pad detector the resolution of the same peak at applied voltage 500V is about (15%), but in case of pixel detector the resolution is worse than in case of pad detector as shown in figure (5b), for Si pixel detector the resolution at 200V is about (8%) and for CdTe pixel detector detector the resolution at 750V is about (10%) and for GaAs pixel detector at 750V is about (24%).

It is possible to operate the detector the pixel detectors with different frequency , Figure 7 (a,b,c), presents the results of operate the detector at two different frequency 10MHz and 25MHz as can seen from the figure there is no difference between operating pixel detector in different frequency for cluster size, signal and resolution, so it is possible to say it is just coefficient magnification of the amplitude (TOT).

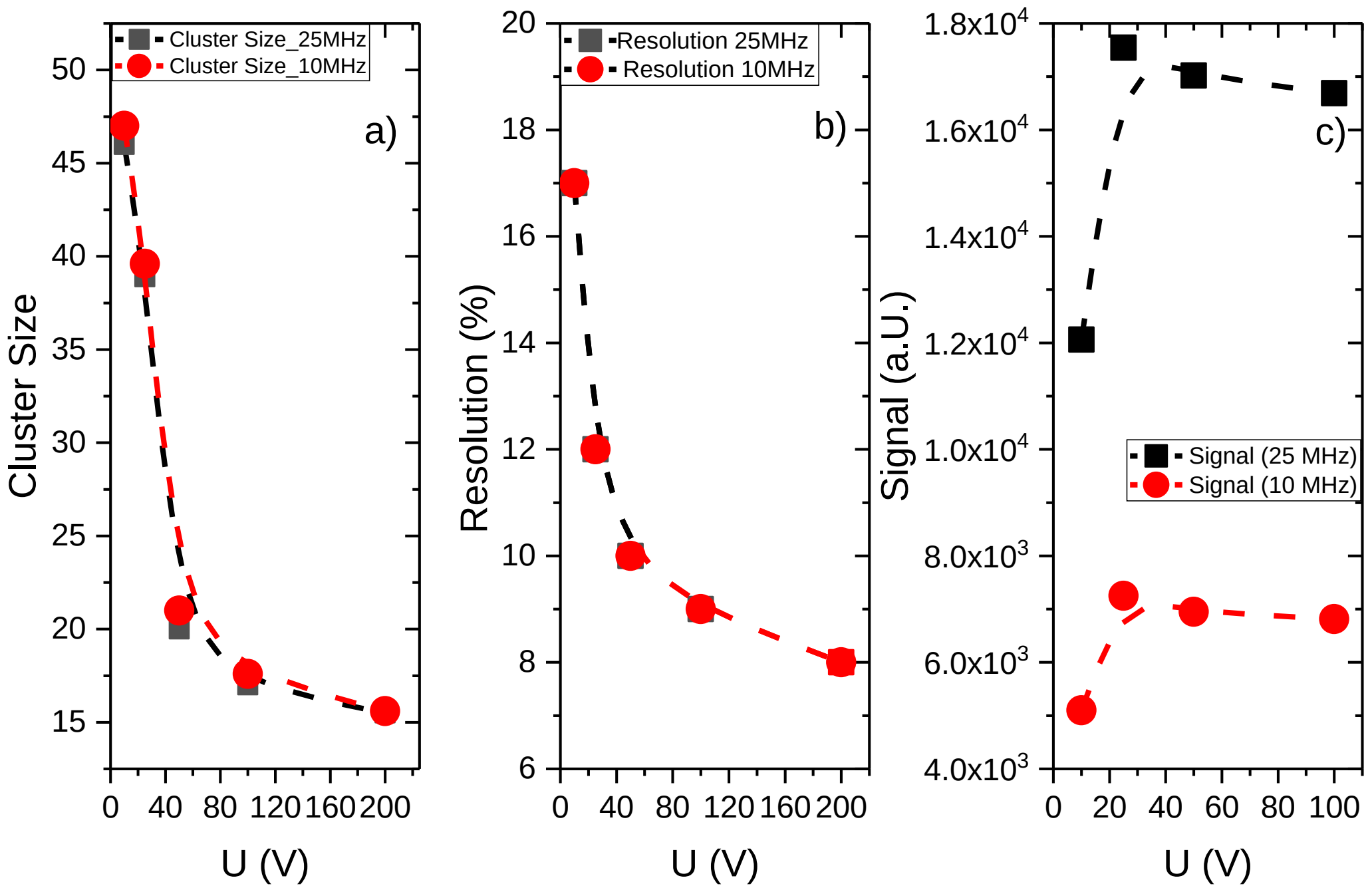

Figure 7 a),b),c) Cluster size, resolution and signal magnitude as a function of applied voltage for different clocks 10MHz and 25MHz for Si pixel detector using 226Ra.

**Section 2 (137Cs):**

In these section focused on the response of pad and pixel detector for 137Cs, Figure (8a) presents spectrum from Si pad detector and Si pixel detector of 137Cs, applied voltage 200 V, as can be seen from the figure, peak from 137Cs are appear in case of Si pad detector and also in pixel detector the peak at (661.6 KeV), the resolution for Si pad detector is about (12%) and for Si pixel detector at the same bias voltage (26%) and may be related to registration efficiency for Si very low at such energy and low thickness of Si pad detector 400µm and 300µm for Si pixel detector. Figure (8b) shows the signal of CdTe, CZT pad detector and CdTe pixel detector of 137Cs, applied voltage 750 V, as can be seen from the figure, peak of 137Cs 661.6 KeV is appear in case of CdTe pad detector and also in CdTe pixel detector , the resolution for CdTe and CZT pad detector (12% and 29%) and for CdTe pixel detector (19%) at the same bias volt. It is clear that in CdTe pad detector and pixel detector the Cs137 peak is very clear because in case CdTe the charge collection both electron and hole but in case of CZT the charge collection only electron , so these may be related to the 137Cs peak not appear. As we mentioned before the detector not calibrated by pixel , our calibration only pre cluster , it is possible to get spectrum using 1,2,3,… pixel count , the resolution becomes better , but the problem that there is no per pixel calibration in such energy, also the type of CZT sensor (nord) was prepared for high energy response. Figure (8c) shows the signal from GaAs pad detector and GaAs:Cr pixel detector[13] by 137Cs, applied voltage 500 V. As can be seen there is no peak (662 keV) in GaAs pad detector. That's because this type of semiconductor has the only electron type charge collection for chromium compensated n-type GaAs:Cr [6]. It is possible only to fit the right border of the spectrum. But in case of pixel detector the peak at (661.6 KeV) are presented, the resolution for GaAs pixel detector is 6% when for GaAs pad detector at the same bias voltage it is about 40%. These may be also related that the thickness of pixel detector GaAs 500µm , and thickness of GaAs pad detector 250µm. For the energy 662keV in such sensors, we have to take in account the low registration efficiency for the used sensor, practically about 0.03 % for GaAs (250 µm). For example the registration efficiency for the more thick CdTe (1000 µm) is about 0.6% [14], from the figure 8 (c) , there is a peak for CdTe and GaAs pixel and pad detector at low energy about 20 keV, these peak related to decay of 137Cs to 137Ba [11].

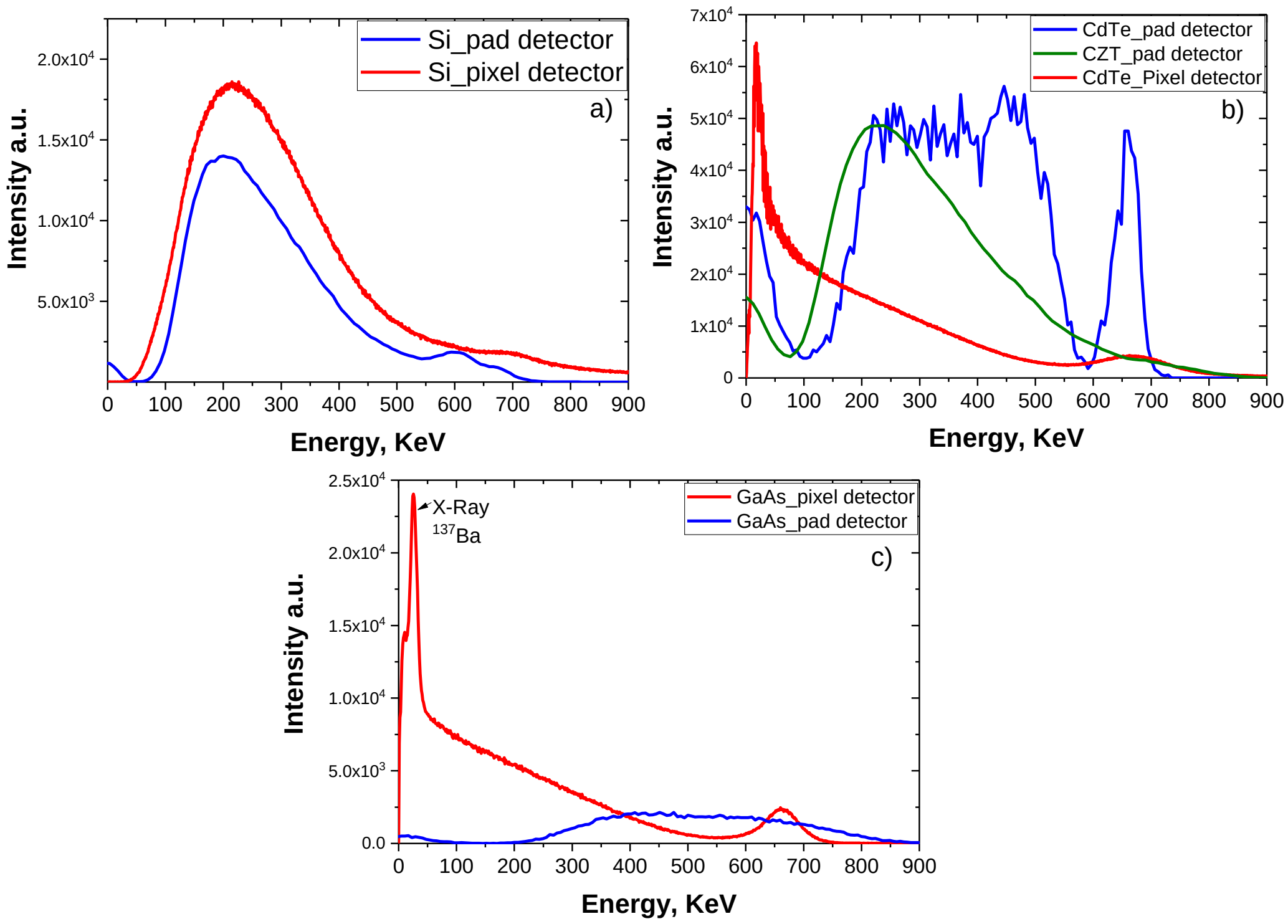

Figure 8 a),b),c) 137Cs spectrum from Si,CdTe,CZT,GaAs:Cr pad and pixel detector.

**Bias voltage effect on charge collection (signal magnitude) and resolution for gamma rays:**

Figures 9 (a,b) presents the relation between the signal and applied volt for pad and pixel detector under 137Cs. The increase of the applied volt leads to increase the signal, then the signal reaches saturation for all pad and pixel detectors. With 137Cs the behavior is same like in case of 226Ra. As mentation before the energy resolution is important parameter for any detector, figure 10(a,b) presented the relation between the resolution and applied volt for pad and pixel detectors. The resolution of 137Cs peak (662 keV) for pad detector becomes better with the increase of applied voltage as expected, up to applied voltage 200 V (fig. (10a). The resolution of Si pad detector is about (10%) at applied voltage 200V and for CdTe up to 750V the resolution is about (10%) and for CZT at 750V the resolution is about (25%), but in case of GaAs the resolution at applied voltage 500V is about (40%), but in case of pixel detectors the best resolution for Si at 200V is about (25%) and for CdTe the resolution at 750V is about (20%) and about (8.9%) for GaAs at 500V.

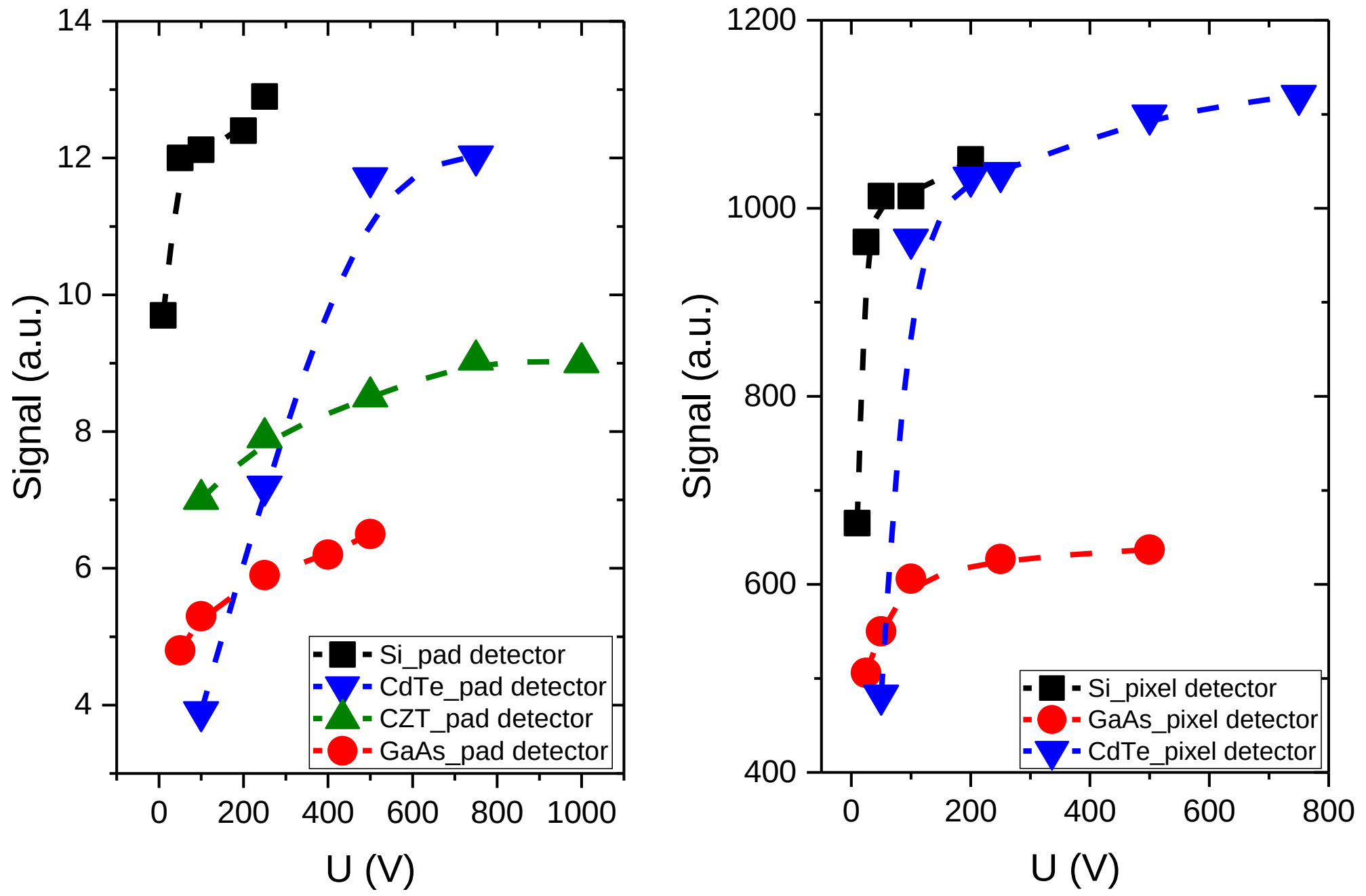

Figure 9 (a,b) Signal from Si,CdTe,CZT,GaAs:Cr pad and pixel detector of 137Cs.

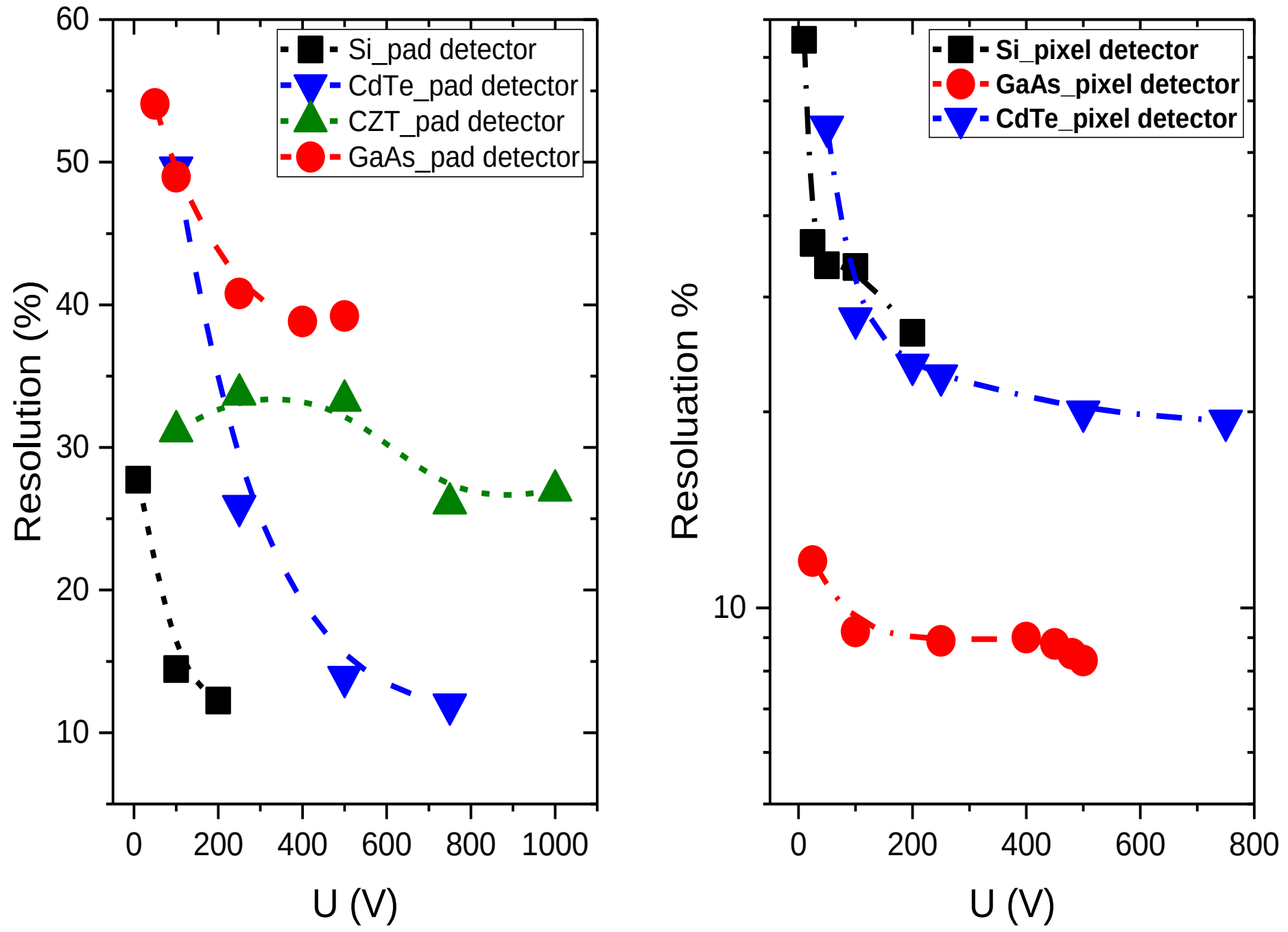

Figure (10) Resolution versus applied voltage (a) pad detector
( b) pixel detector using 137Cs

**Conclusions:**

The bias voltage, as expected, has a large impact on the detector performance. The cluster size and resolution depends on the bias voltage due to two competing effects: less diffusion and better charge collection, which are the results of increasing the bias voltage. Other conclusions are that at higher bias voltage the cluster size decreases and improve the resolution. For single-pad detectors the γ- and α- radiation showed that to register high energy γ it is preferable to use CdTe due to its high efficiency and good, the 662 keV γ- peak of 137Cs stands out well. GaAs:Cr and CdZnTe cannot be used to separate γ peaks because these detectors collect only the electron component of the charge. For 4-7 MeV α-spectra all 4 (Si, GaAs:Cr, CdTe, CdZnTe) are acceptable, Si has the best resolution, but GaAs:Cr can't separate the 5.30 MeV and 5.49 MeV peaks. The γ and α radiation response for Timepix3 pixel detectors showed that the resolution of pixel detectors is generally worse than that of single pad detectors. This can be explained that the charge is divided to several pixels, which leads to the accumulation of errors, and pixel-by-pixel calibration was not performed. CdTe and GaAs pixel detectors are preferred for high energy γ detection due to their higher efficiency than Si and still good resolution for the 660 keV γ of 137Cs. Alpha spectra in pixel detectors showed that Si and CdTe detectors are preferable, with Si having the best resolution, while GaAs:Cr could only resolve the peak 7.7 MeV while combining peaks at 4.8, 5.4 and 6.0 MeV.